\documentclass[%
 reprint,
superscriptaddress,
nofootinbib,
 amsmath,amssymb,
 aps,
prb,
]{revtex4-1}
\usepackage{graphicx}% Include figure files
\usepackage{dcolumn}% Align table columns on decimal point
\usepackage{bm}% bold math
\usepackage{xcolor}
\usepackage{subfigure}
\usepackage{textcomp}
\makeatletter
\newcommand{\mathleft}{\@fleqntrue\@mathmargin0pt}
\newcommand{\mathcenter}{\@fleqnfalse}
\makeatother
\begin{document}

\preprint{APS/123-QED}

%referees: tolias,bostrom,bergstrom,kimbal,palasantzas

\title{Force-dependent elasticity of nucleic acids}%

\author{Juan Luengo-M\'arquez}
\email{juan.luengo@uam.es}
\affiliation{Departamento de Física Teórica de la Materia Condensada, Universidad Autónoma de Madrid, 28049 Madrid (Spain)}
\affiliation{Instituto Nicol\'as Cabrera, Universidad Autónoma de Madrid, 28049 Madrid (Spain)}
\author{Juan Zalvide-Pombo}%
%\email{email2@uam.es}
\affiliation{Departamento de Física Teórica de la Materia Condensada, Universidad Autónoma de Madrid, 28049 Madrid (Spain)}
\author{Rub\'en P\'erez}%
%\email{email3@uam.es}
\affiliation{Departamento de Física Teórica de la Materia Condensada, Universidad Autónoma de Madrid, 28049 Madrid (Spain)}
\affiliation{Instituto Nicol\'as Cabrera, Universidad Autónoma de Madrid, 28049 Madrid (Spain)}
\affiliation{Condensed Matter Physics Center (IFIMAC), Universidad Autónoma de Madrid, 28049 Madrid (Spain)}
\author{Salvatore Assenza}%
\email{salvatore.assenza@uam.es}
\affiliation{Departamento de Física Teórica de la Materia Condensada, Universidad Autónoma de Madrid, 28049 Madrid (Spain)}
\affiliation{Instituto Nicol\'as Cabrera, Universidad Autónoma de Madrid, 28049 Madrid (Spain)}
\affiliation{Condensed Matter Physics Center (IFIMAC), Universidad Autónoma de Madrid, 28049 Madrid (Spain)}


%\affiliation{
%\affiliation{
%\affiliation{%


\date{\today}% It is always \today, today,

\begin{abstract}  
	The functioning of double-stranded (ds) nucleic acids (NAs) in cellular processes is strongly mediated by their elastic response. These processes involve proteins that interact with dsDNA or dsRNA and distort their structures. The perturbation of the elasticity of NAs arising from these deformations is not properly considered by most theoretical frameworks. In this work, we introduce a novel method to assess the impact of mechanical stress on the elastic response of dsDNA and dsRNA through the analysis of the fluctuations of the double helix. Application of this approach to atomistic simulations reveals qualitative differences in the force dependence of the mechanical properties of dsDNA with respect to those of dsRNA, which we relate to structural features of these molecules by means of physically-sound minimalistic models.
\end{abstract}

\maketitle

There is an intricate connection between the elasticity of nucleic acids and their biological role\cite{hogan1987,rohs2009,rohs2010,da2021,sponer2018,yi2022,vicens2022,marin2021}, which affects their organization and functionality over multiple scales\cite{smith1996,herrero2013,fosado2016,mdp2016,mdp2017,sponer2018,vicens2022}. Specific motifs, such as A-tracts in double-stranded DNA (dsDNA) and AU-tracts in double-stranded RNA (dsRNA), have been found to confer to the chain mechanical properties significantly departing from sequence-averaged values\cite{moreno2006,segal2009,haran2009,marin2020,drvsata2014,marin2020b}, supporting the existence of a mechanical code contained in the sequence of nucleotides along with the genetic code\cite{eslami2016,marin2019}.
\vspace{0.10cm}
\\
When stretching forces in the range 5-50 pN are exerted, as is the case e.g. for RNA polymerases\cite{yin1995}, the mechanical response of double-stranded nucleic acids lies within the elastic regime\cite{smith1996,baumann1997,marko1994}, where enthalpic structural deformations dominate while the molecules still retain the double-helical conformation\cite{noy2012,marin2021}. The mechanical properties of nucleic acids are theoretically depicted by mapping the set of conformations of the molecule into a set of energy values by means of a suitable functional form. For deformations close to the equilibrium conformation of the chain\cite{nelson2004}, the mechanical energy is approximately harmonic in the deformation modes. At the simplest level, double-stranded nucleic acids can be regarded as homogeneous elastic rods, characterized by their extension and torsion. The Elastic Rod Model (ERM) energy of the system upon pulling thus reads\cite{gore2006}
\begin{equation}
	E(\Delta L,\Delta\theta) = \frac{1}{2}\frac{S}{L_{0}}\Delta L^{2} + \frac{1}{2}\frac{C}{L_{0}}\Delta \theta^{2} + \frac{g}{L_{0}}\Delta L\Delta\theta - \Delta L f
	\label{eq:elastic_rod_model}
\end{equation}
where $\Delta L$ and $\Delta \theta$ are the contour length and twist deformations with respect to their equilibrium values, $L_{0}$ is the equilibrium contour length of the molecule, and the stretch modulus $S$, the twist modulus $C$ and the twist-stretch coupling $g$ are the elastic parameters. In Eq. \ref{eq:elastic_rod_model} the thermal bending has been neglected, which is a reasonable assumption for chains significantly shorter than the persistence length, as the ones studied here.
\vspace{0.10cm}
\\
Substantial experimental and theoretical efforts have been made to characterize the elastic response of nucleic acids\cite{bustamante2003,herrero2013,orozco2003,orozco2008,marin2021,gore2006,marko1997,lipfert2014,liebl2015,marin2017}, revealing a number of striking mechanical features of dsDNA, such as the positive correlation between the deformations in the contour length and the torsion angle for forces up to approximately 40 pN\cite{marko1997,gore2006}. Conversely, dsRNA has been found to unwind when stretched\cite{lipfert2014,liebl2015,marin2017}, which is the typical behavior of most chiral materials\cite{landau1986vol7}.
\vspace{0.10cm}
\\
The picture becomes even more complex when one realizes that the elastic parameters of the ERM depend themselves on the mechanical stresses exerted on the molecule\cite{gore2006,gross2011,shi2013,broekmans2016,marin2017}. For instance, rotor-bead tracking has shown that the change in torsion of pulled dsDNA becomes negative beyond 40 pN\cite{gore2006}, which highlights a change in sign for $g$. This evinces a strong limitation in the description of nucleic acids by means of the ERM with constant parameters\cite{mathew2008}, as this model cannot account for the change in the elastic response produced by the microscopic distortion of the chain. To address the force-dependent mechanics, a subsequent work assumed a force-dependent twist-stretch coupling, $g(f)$, and determined empirically that $g$ was roughly constant up to 30 pN, after which it increased linearly with the pulling force\cite{gross2011}. However, this approach assumes that the whole force-dependence must rely on the twist-stretch coupling, while it would be equally expectable to find a force-dependence on any of the elastic parameters. Moreover, the quantitative details of $g(f)$ depend on the values chosen for $S$, $C$ and the persistence length $l_{P}$, often leading to physically-unacceptable imaginary values of $g(f)$ even within the experimentally-observed ranges of values for the three constants (Section S1 in the \textit{Supplemental Material}). 
\vspace{0.10cm}
\\
Current theoretical approaches are ill-suited to extract the elastic parameters with force-dependence. On the one hand, the standard practice of fitting the average values by means of formulas derived from the minimization of Eq. \ref{eq:elastic_rod_model} cannot account by construction for stress-dependent elastic constants, since a stress-strain curve is needed for fitting purposes. Interestingly, for a monotonous change of e.g. $S(f)$, the effective value obtained from fitting lies outside of the range of values spanned by $S(f)$ within the force domain (Section S2 \textit{Supplementary Material}), thus introducing a systematic offset in the estimation of the parameter. On the other hand, fluctuations-based approaches\cite{go1976,olson1998} are widely used\cite{noy2012,velasco2020,drvsata2014}, but are based on formulas strictly valid only in the absence of external mechanical stress.
\vspace{0.10cm}
\\
In this Letter, we present an alternative route to explore the force dependence of the parameters of the ERM based on a novel generalization of the latter approach, which enables the study of fluctuations in the presence of mechanical stresses. Application of our method to atomistic trajectories of dsDNA and dsRNA sequences from literature\cite{marin2017,marin2019,marin2019b} unveils a hitherto unreported dependence of the stretch and twist moduli on the stretching force, whose physical origin is identified by means of minimalistic toy models. These novel predictions on $S(f)$ and $C(f)$ are found together with a behavior of $g(f)$ in line with experimental observations.
\vspace{0.10cm}
\\
The ERM belongs to a class of models attributing a harmonic energetic penalty for each of the deformation modes, including also coupling terms between different modes\cite{olson1998}. We denote as $\Delta q_{i}$ the deformation associated to the generalized coordinate $q_{i}$, and as $\gamma_{i}$ the generalized force conjugated to $q_{i}$. The energy for $N$ deformation modes thus reads
\begin{equation}
	E(\boldsymbol{\bar{q}}) = \frac{1}{2}\sum_{i}^{N}\sum_{j}^{N}k_{ij}\Delta q_{i}\Delta q_{j} - \sum_{i}\gamma_{i}\Delta q_{i}
	\label{eq:generalized_energy}
\end{equation}
where $k_{ii}$ is the elastic modulus of the mode $i$, and $\frac{1}{2}(k_{ij} + k_{ji})$ for $i\neq j$ is the coupling factor between the modes $i$ and $j$.
\vspace{0.10cm}
\\
We first define the work matrix, having elements $\left[\boldsymbol{\bar{\bar{\Gamma}}}\right]_{ij} = \left< \Delta q_{i} \right> \gamma_{j}$, and the covariance matrix with elements $\left[\boldsymbol{\bar{\bar{V}}}\right]_{ij} = \left< \Delta q_{i} \Delta q_{j} \right>$. From Eq. \ref{eq:generalized_energy} and the generalized equipartition theorem\cite{huang2008}, $\left<\Delta q_{i}\frac{\partial E}{\partial q_j}\right>=\delta_{ij} k_{B}T$, it is possible to show that (see Section S3 in \textit{Supplementary Material})
\begin{equation}
	\boldsymbol{\bar{\Bar{K}}} = \left(\boldsymbol{\bar{\bar{V}}}\right)^{-1}\left(k_{B}T\boldsymbol{\bar{\bar{I}}} + \boldsymbol{\bar{\bar{\Gamma}}}\right)
        \label{eq:force-dependent_fluctuations}
\end{equation}
where the generalized stiffness matrix has elements $\left[\boldsymbol{\bar{\bar{K}}}\right]_{ij} = \frac{1}{2}(k_{ij} + k_{ji})$, $k_{B}T$ is the thermal energy of the system, and $\boldsymbol{\bar{\bar{I}}}$ is the $N$th order identity matrix. Note that the diagonal elements of the stiffness matrix are the elastic moduli, and the off-diagonal elements are the coupling parameters. Remarkably, Eq. \ref{eq:force-dependent_fluctuations} establishes a method to compute rigorously the parameters of the ERM from the knowledge of the fluctuations associated to the deformation modes of a perturbed system. Additionally, since $\boldsymbol{\bar{\bar{I}}}$ is symmetric, Eq. \ref{eq:force-dependent_fluctuations} imposes that any system described by Eq. \ref{eq:generalized_energy} must satisfy $\left[\boldsymbol{\bar{\bar{V}}}\boldsymbol{\bar{\bar{K}}}-\boldsymbol{\bar{\bar{\Gamma}}}\right]_{ij} = \left[\boldsymbol{\bar{\bar{V}}}\boldsymbol{\bar{\bar{K}}}-\boldsymbol{\bar{\bar{\Gamma}}}\right]_{ji}$. The fulfillment of such requirement allows the evaluation of the extent to which a physical system may be correctly described by the harmonic approximation.
\vspace{0.10cm}
\\
We next exploit this approach by investigating the elastic properties of short dsDNA and dsRNA molecules when a stretching force up to 20 pN is applied, for which the stretched ERM energy with two deformation modes (Eq. \ref{eq:elastic_rod_model}) should be a sufficient description of the elastic response. We have analyzed the trajectories of 17 dsDNA sequences and 11 dsRNA sequences from all-atoms simulations reported in the literature\cite{marin2017,marin2019,marin2019b}, based on the parm99 force field\cite{wang2000} with the bsc0 modifications \cite{perez2007} and, in the case of dsRNA, the $\chi_{OL3}$ modifications \cite{zgarbova2011}. This force field has been shown to quantitatively account for the different elastic behavior of dsDNA and dsRNA\cite{marin2017} and has been employed to inform a coarse-grained model capable of quantitatively recapitulating the experimental findings on dsDNA elasticity\cite{assenza2022}.  The set of sequences can be found in Tables S1 and S2, while the force-dependent elastic parameters obtained for each sequence by means of Eq. \ref{eq:force-dependent_fluctuations} can be found in Section S5 in the \textit{Supplementary Material}.
\vspace{0.10cm}
\\
To convey succinctly the information contained in our analysis, we computed the variations $C'(f)$, $S'(f)$ and $g'(f)$ of the elastic constants. Quantitatively, we considered the slopes of linear fits of their force dependence, which account for the sign and magnitude of such variation. This approach implicitly assumes a linear dependence, which holds acceptably for most sequences (see Fig. S9-S36). The appropriateness of the linear approximation is accounted for in the error estimation of the slope.
\vspace{0.10cm}
\\
\begin{figure}[ht!]
\includegraphics[width=0.42\textwidth,height=0.59\textwidth,keepaspectratio]{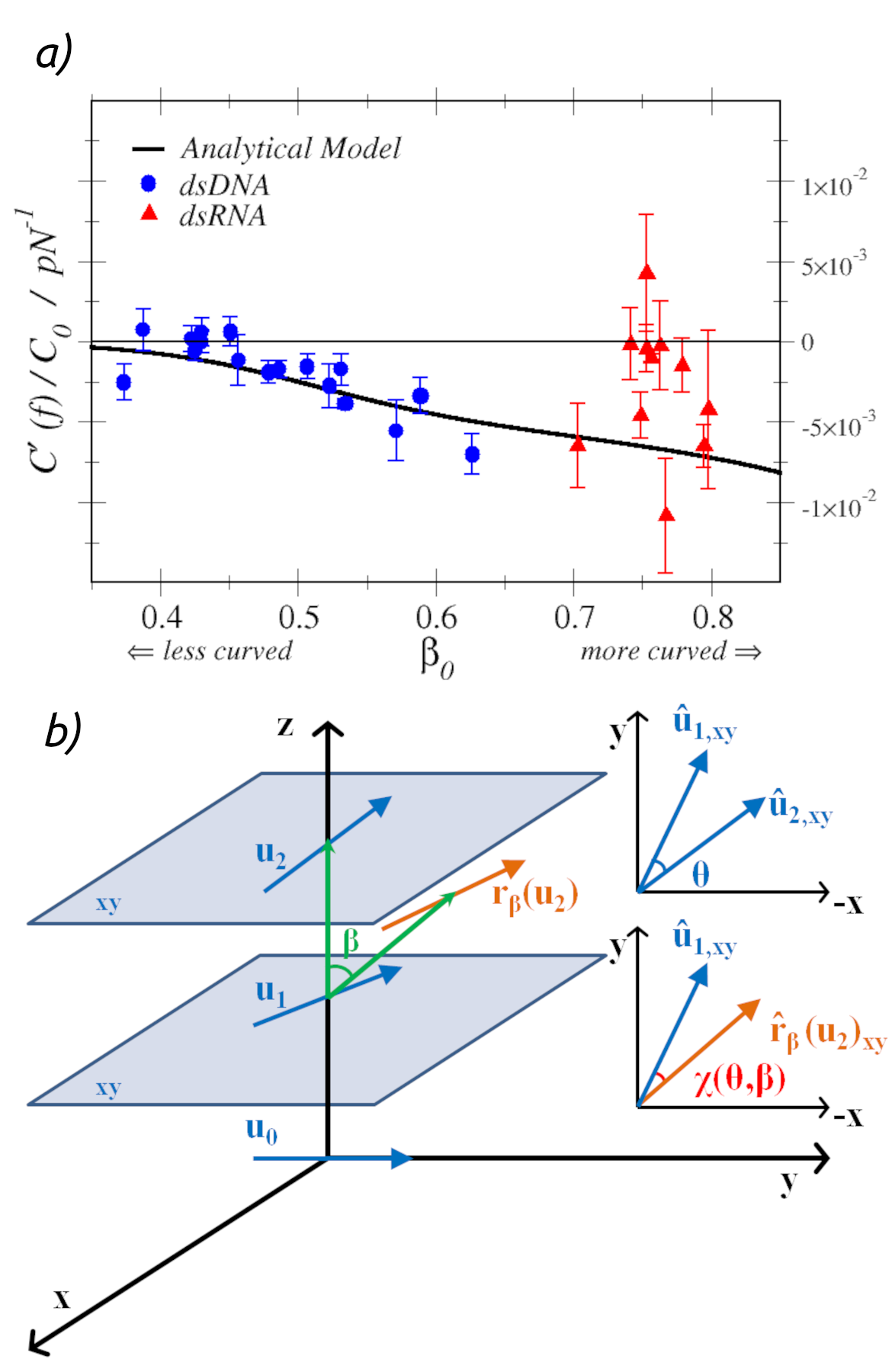}
	\caption{
		(a) Relative variation of the twist modulus with the force against the crookedness at zero force. For the scatter plot, $\beta_{0}$ and $C_{0}$ indicate the crookedness and the twist modulus at $f=1\ pN$. (b) Schematic representation of the toy model.}
\label{fig:twist}
\end{figure}
For the twist modulus we find $C'(f)\leq0$ within error for all sequences. Furthermore, in the case of dsDNA, $C'(f)$ displays a strong negative correlation with the crookedness $\beta$ (Fig. \ref{fig:twist} a), which quantifies the displacement of the centers of the base pairs from the helical axis\cite{marin2019}. To account for this displacement, as $\beta$ increases, the normals to the base planes must be less aligned with the helical axis. Since the local torsion is mostly determined by the stacking between bases, while the overall twist modulus is obtained by the change in the helical twist, we conjecture that the change in $C$ is due to the progressive alignment of the base pairs and the helical axis imposed by the force.
\vspace{0.10cm}
\\
In order to test this hypothesis, we develop a toy model with the minimal components that illustrate this concept (see Fig. \ref{fig:twist} (b)). We place three vectors, $u_{0}$, $u_{1}$ and $u_{2}$, along the z-axis, which is oriented parallel to the helical axis.  Each vector is pointing from one base to the opposite one at a certain basepair of a perfectly straight molecule, i. e. $u_{0}$, $u_{1}$ and $u_{2}$ are all perpendicular to the $z$ axis. The twist angle obtained in this case is the intrinsic torsion angle $\theta=\arccos(u_{1}\cdot u_{2})$. To account for the presence of a spontaneous curvature, we rotate the top vector ($u_{2}$) by an angle $\beta$ around the x axis, obtaining the rotated vector $r_{\beta}(u_{2}$). To consider a crookedness for this simplified geometry, one can think of the centers of the three vectors as being extracted from a helix with helix angle $\beta$, for which a direct identification with the crookedness can be shown formally (see Section S4.2 in the \textit{Supplementary Material}). The torsion angle $\chi(\theta,\beta)$ is then the angle formed by the versors of $u_{1}$ and of the projection of $r_{\beta}$ onto the $xy$ plane.
\vspace{0.10cm}
\\
Next, we assume that the energetics of the torsion is determined by the intrinsic torsion angle as $E(\theta)=\frac{\widetilde{C}}{2L_{0}}(\theta-\theta_0)^2$, where  the intrinsic twist modulus $\widetilde{C}$ is independent of the force and $\theta_{0}$ is the equilibrium twist angle. By the equipartition theorem, $\widetilde{C}=\frac{L_{0}k_{B}T}{\left<\Delta \theta^2\right>}$. In contrast, the observed twist modulus is obtained as $C(\beta)=\frac{L_{0}k_{B}T}{\left<\Delta \chi^2\right>}$, where the variance $<\Delta \chi^2>$ can be computed by standard Boltzmann statistics considering the functional dependence of $\chi$ on $\theta$ and $\beta$ (see Section S4.3 in the \textit{Supplemental Material}). When $\widetilde{C}/L_{0}k_{B}T\ \gg 1$, the twist modulus may be expanded up to
\begin{equation}
	C(\beta) = C(\beta_{0})\left[1 + \sin(\beta_{0})\Omega(\theta_{0},\beta_{0})\left(\beta - \beta_{0}\right)\right],
	\label{eq:twist_modulus}
\end{equation}
being $\Omega(\theta_{0},\beta_{0})$ a function solely evaluated at the equilibrium values of the intrinsic torsion angle and bending angle, and whose analytic functional form is presented in Section S4.3 in \textit{Supplementary Material}. Making use of Eq. \ref{eq:twist_modulus}, we compute $C'(f) = \frac{\partial C}{\partial \beta} \cdot \frac{\partial \beta}{\partial\cos(\beta)}\cdot\frac{\partial\cos(\beta)}{\partial f}$. Note also that $\frac{\partial\cos(\beta)}{\partial f} = \cos(\beta)/k_{\beta}$, where $k_{\beta}$ is the stiffness associated to the crookedness\cite{marin2019}. Considering that to first order $\sin(\beta)\approx\sin(\beta_{0})$, we can thus write
\begin{equation}
	\frac{C'(f)}{C(\beta_{0})} = -\frac{\cos(\beta_{0})\Omega(\theta_{0},\beta_{0})}{k_{\beta}(\beta_{0})}
	\label{eq:toy_model_result}
\end{equation}
In order to make use of this result, we employ the fit performed by Marin-Gonzalez \textit{et al.}\cite{marin2019}, in which $k_{\beta}$ could be expressed as a function of $\beta_{0}$ as $k_{\beta}(\beta_{0}) = Ae^{-k\beta_{0}} + B$, being $A = (2.24 \pm 1.24)\times 10^{6} \ pN$, $B = 700 \pm 120 \ pN$, and $k = 16.2 \pm 1.5$. We show in Fig. S5 that the previous fit captures the $k_{\beta}(\beta_{0})$ function for all sequences analyzed in this work, including those of dsRNA.
\vspace{0.10cm}
\\
The analytical result in Eq. \ref{eq:toy_model_result} conveys the variation of the twist modulus with the stretching force, employing as only input the equilibrium intrinsic torsion angle and the equilibrium crookedness. In Fig. \ref{fig:twist} (a) we display the values of $C'(f)/C(\beta_{0})$ from the analysis of the simulations, identifying $\beta(f = 1 pN)$ with $\beta_{0}$. We also show the analytical result of Eq. \ref{eq:toy_model_result} taking the equilibrium intrinsic torsion angle of dsDNA $\theta_{0} = 34^{\circ}$. Despite being minimalistic, the toy model provides a prediction (Eq. \ref{eq:toy_model_result}) that captures quantitatively the data obtained for dsDNA without any fitting (Fig. \ref{fig:twist} (a)). Indeed, as discussed above, the parameters $\theta_{0}$ and $k_{\beta}$ in Eq. \ref{eq:toy_model_result} were set \textit{a priori} to established values. The perturbative approach used to reach Eq. \ref{eq:toy_model_result} is justified considering that for dsDNA $\widetilde{C}/L_{0}k_{B}T\ \approx 35$. For comparison, in Fig. S6 we also show the curves obtained numerically for dsDNA and dsRNA without the perturbative approximation. The toy model reveals that the observed decrease of the twist modulus with the stretching force can be explained solely by geometrical arguments, being a direct consequence of the intrinsic crookedness of the molecules.
\begin{figure}[htb!]
\includegraphics[width=0.42\textwidth,height=0.59\textwidth,keepaspectratio]{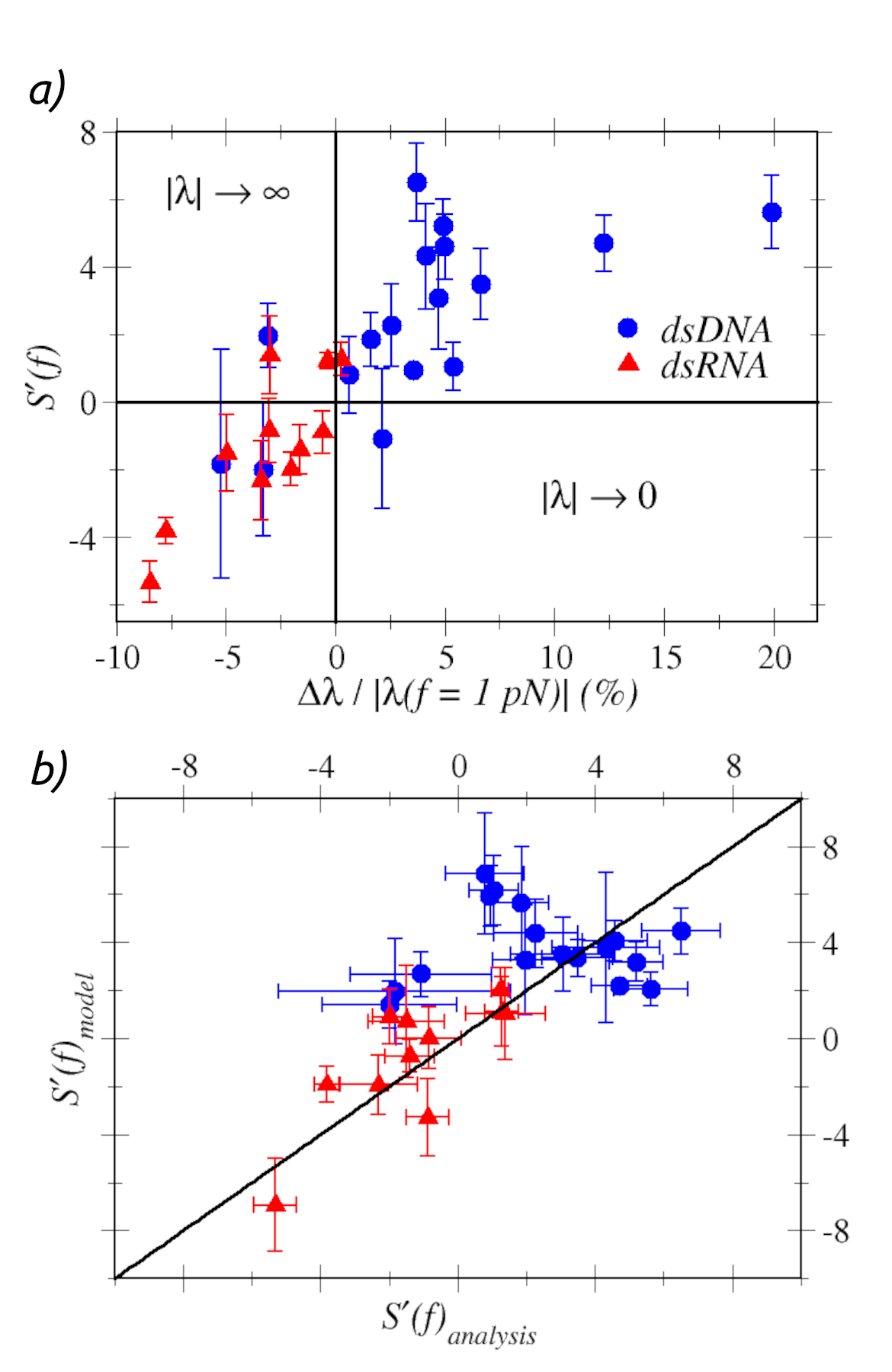}
	\caption{(a - top) Variation of the stretch modulus with the force against the relative slide variation. (b - bottom) Scatter of $S'(f)$ comparing the predictions of the model in Ref.\cite{marin2019} and our analysis.}
\label{fig:stretch}
\end{figure}
\vspace{0.10cm}
\\
In the case of dsRNA, the simulation data do not show a clear correlation between $C'(f)/C(\beta_{0})$ and $\beta_{0}$ (Fig. \ref{fig:twist} (a)), suggesting that at large curvatures the crookedness might not be the key determinant of the twist response of the molecule. A possible concurring factor might be, for instance, a force-dependent intrinsic twist modulus $\widetilde{C}$. Despite not capturing quantitatively the data, the toy model does however account for the sign and order of magnitude of $C'(f)/C(\beta_{0})$.  
\vspace{0.10cm}
\\
We now focus on the stretch modulus $S$. As shown in Fig. \ref{fig:stretch} (a), we first observe that for most dsDNA sequences the stretch modulus increases, $S'(f)>0$, while for most dsRNA sequences $S'(f)<0$. A possible microscopic mechanism to which this observation may be ascribed is the strengthening - or weakening - of the stacking interactions between the bases upon stretching. Such effect can be monitored by analyzing the force-variation of the slide, $\lambda$, a structural parameter that measures the relative displacement of two consecutive base pairs along the direction of the Watson-Crick hydrogen bonds. The slide is known to have a different evolution with the stretching force for dsDNA and dsRNA\cite{marin2017}.
\vspace{0.10cm}
\\
Fig. \ref{fig:stretch} (a) displays $S'(f)$ against the relative slide variation of each sequence $\Delta\lambda/|\lambda(f = 1\ pN)|$, with $\Delta\lambda = \lambda(f = 20\ pN)-\lambda(f = 1\ pN)$. 
Since $\lambda < 0$, a negative slide variation - as found in most dsRNA sequences - implies that $\lambda$ increases in magnitude, while $\Delta \lambda > 0$  - as found in most dsDNA sequences - implies $|\lambda|\rightarrow 0$. We reason that the magnitude of the slide is expected to be negatively correlated with the strength of the stacking interactions, since $|\lambda|$  quantifies the degree at which the aromatic rings in the stacking bases are not overlapping with each other. Since the slide is negative for both dsDNA and dsRNA, the points in Fig. \ref{fig:stretch} (a) corresponding to $\Delta\lambda/|\lambda(f = 1\ pN)|<0$ indicate that the force further reduces the overlap of bases, i.e. it weakens the stacking interactions, hence $S'(f)<0$. Conversely, $\Delta\lambda/|\lambda(f = 1\ pN)|>0$ is indicative of a strengthening of the stacking interactions, so that $S'(f)>0$. Microscopically, the opposite change of $\lambda$ for stretched dsDNA and dsRNA can be traced down to the different way in which it correlates with the h-rise for the two molecules\cite{marin2017}. As we discuss in Section S4.5 in the \textit{Supplemental Material}, in dsDNA the magnitude of the slide is negatively correlated with the h-rise, hence the increase of the h-rise upon pulling results in a decrease of $|\lambda|$. In contrast, dsRNA is characterized by a positive correlation, so that the slide further increases in magnitude upon pulling.
\vspace{0.10cm}
\\
These data can be further interpreted in terms of the model proposed by Marin-Gonzalez \textit{et al.} in Ref.\cite{marin2019}, where the stretch modulus was expressed as the effective spring constant of a harmonic summation of base-pair steps stiffnesses $k_{bp}$ and the crookedness stiffness, $\frac{1}{S}=\frac{1}{k_{\beta}}+\sum\frac{1}{k_{bp}}$. Assuming $k_{bp}$ not to change upon pulling, one thus finds $S'(f)=\left[\frac{S}{k_{\beta}}\right]^{2}k_{\beta}'(f)$. Hence, the sign of the variation of the stretch modulus is determined by the sign of $k_{\beta}'(f)$. The crookedness stiffness estimates the energy cost of reducing the crooked curvature of the molecule, and the stronger the stacking interactions, the larger this energy cost. Fig. \ref{fig:stretch} (b) shows how the predictions of the model match the sign and order of magnitude of $S'(f)$ from the analysis of fluctuations, thus providing a plausible physical interpretation of the results.
\vspace{0.10cm}
\\
Finally, for the variation of the twist-stretch coupling, we find $g'(f)\geq0$ for virtually all dsDNA and dsRNA sequences (Fig. S10). This observation matches the experiments for dsDNA\cite{gore2006,gross2011}, where it was found that $g$ should switch to a positive sign at approximately 40 pN. In the case of dsRNA, ${g'(f)>0}$ implies an enhancement of the negative correlation between twist and stretch. Ordinary chiral objects are expected to have $g(f)>0$\cite{kamien1997}. In Section S4.7 in \textit{Supplementary Material} we show that a helical object with constant radius has $g'(f)<0$, so that the observation of $g'(f)>0$ is far from being trivial. It would be interesting to compare this prediction based on all-atom simulations to experimental data. For instance, in a rotor-bead assay the combination $g(f)>0$ and $g'(f)>0$ would result in a non-linear decrease of torsion with force with negative convexity. To our knowledge, this feature has never been explored experimentally. 
\vspace{0.10cm}
\\
In this study, we have introduced a novel approach that rigorously allows the computation of the stress-dependent elastic constants of the generalized Elastic Rod Model for molecules under the action of external work. We believe that this procedure sets a promising theoretical playground for future works, opening the field to overcome the limitations of the ERM, whose requirement of constant mechanical moduli hampers the full characterization of mechanical phenomena emerging from the complexity of nucleic acids. Our analysis of atomistic simulations reveals that the twist modulus decreases upon stretching for both dsDNA and dsRNA. We show that this variation arises from the intrinsic curvature of the molecules, and that a toy model depicting this argument can account quantitatively for the simulation data in the case of dsDNA, while capturing the qualitative behavior of dsRNA. Most strikingly, we find that the stretch modulus of dsDNA becomes stiffer upon pulling, while that of dsRNA softens. Based on our analysis, we ascribe this difference to the opposite force-response of stacking interactions in dsDNA and dsRNA. Finally, the twist-stretch coupling is found to increase with force for both dsDNA and dsRNA. This is in line with the experimental observations for dsDNA, while it gives a novel prediction for dsRNA to be tested in single-molecule setups. Our study establishes new distinctive features of nucleic acids, further enlarging the list of fundamental differences between dsDNA and dsRNA.
\vspace{0.10cm}
\\
The project that gave rise to these results received the support of a fellowship from “la Caixa” Foundation (ID 100010434) and from the European Union’s Horizon research and innovation programme under the Marie Skłodowska-Curie grant agreement No. 847648. The fellowship code is LCF/BQ/PI20/11760019. We acknowledge support from the Ministerio de Ciencia e Innovaci\'on (MICINN) through the project PID2020-115864RB-I00 and the ``Mar\'ia de Maeztu'' Programme for Units of Excellence in R\&D (grant No. CEX2018-000805-M).

\bibliography{./biblio}% Produces the bibliography via BibTeX.

\end{document}